%% file: eprint.tex
\documentclass[12pt]{article}
\usepackage{graphicx}

\newcommand\pubnumber{
IPPP/20/85, KA-TP-04-2021, P3H-21-018}
\newcommand\pubdate{\today}

\def\instituteT{Institut f\"ur Theoretische Physik, Karlsruhe Institut f\"ur Technologie, 76128 Karlsruhe, Germany}
\def\instituteS{Durham U., IPPP (DH1 3LE, UK) and Oxford U., Theor. Phys. (OX1 3PU, UK)}
\def\supportT{\footnote{The work of T.J. was supported by the DFG under grant  396021762 - TRR 257.}}
\def\supportS{\footnote{UK Science and Technology
Facilities Council (grant number ST/P001246/1)}}

\def\Title#1{\begin{center} {\Large #1 } \end{center}}
\def\Author#1{\begin{center}{ \sc #1} \end{center}}
\def\Address#1{\begin{center}{ \it #1} \end{center}}

\newcommand\pubblock{\rightline{\begin{tabular}{l} \pubnumber\\
         \pubdate  \end{tabular}}}
\newenvironment{Abstract}{\begin{quotation}  }{\end{quotation}}
\newenvironment{Presented}{\begin{quotation} \begin{center} 
             PRESENTED AT\end{center}\bigskip 
      \begin{center}\begin{large}}{\end{large}\end{center} \end{quotation}}
\def\Acknowledgements{\bigskip  \bigskip \begin{center} \begin{large}
             \bf ACKNOWLEDGEMENTS \end{large}\end{center}}
\input econfmacros.tex
\begin{document}
\begin{titlepage}
\pubblock

\vfill
\Title{Non-resonant and offshell effects in top-pair observables}
\vfill
\Author{ Silvia Ferrario Ravasio\supportS}
\Address{\instituteS}
\Author{ Tom\'a\v{s} Je\v{z}o\supportT}
\Address{\instituteT}
\vfill
\begin{Abstract}
In these proceedings we investigate the impact of offshell and non-resonant contributions to kinematic distributions in $t\bar{t}$ final states by comparing several next-to-leading-order~(NLO) + parton shower (PS) generators which differ in the way they model these effects. 
We focus on distributions traditionally used for measurements of fundamental top-quark properties such as its Yukawa coupling or its mass.
Given the current experimental uncertainties, we find that such effects can have relatively large impact on the shape of these distributions and are thus mandatory for accurate measurements.
\end{Abstract}
\vfill
\begin{Presented}
$13^\mathrm{th}$ International Workshop on Top Quark Physics\\
Durham, UK (videoconference), 14--18 September, 2020
\end{Presented}
\vfill
\end{titlepage}
\def\thefootnote{\fnsymbol{footnote}}
\setcounter{footnote}{0}

\section{Introduction}
The top quark plays an important role in the electroweak (EW) symmetry breaking mechanism due to its large coupling to the Higgs boson. 
The precise measurement of this Yukawa coupling, $Y_t$, is an important milestone in the Large Hadron Collider (LHC) physics program.
It has already been measured both directly in $t\bar{t}H$ process \cite{Aaboud:2018urx, Sirunyan:2018hoz} and indirectly in top-pair production \cite{Sirunyan:2019nlw, Sirunyan:2020eds}, in which $Y_t$ appears through EW corrections.

The most recent analysis of Ref.~\cite{Sirunyan:2020eds} relies on 140 fb${}^{-1}$ of top-pair data in the all-leptons channel and on the distributions of the invariant mass of the $bl^+\bar{b}l^-$ system and of the relative rapidity of $bl^+$ and $\bar{b}l^-$.
The measurement yields a best fit value of $Y_t = 1.16^{+0.24}_{-0.35}$.
Variations of the Yukawa coupling in this analysis from $Y_t=1$ to $Y_t=2$ roughly correspond to a shape effect from $+4\%$ to $-2\%$ from lower to higher tails of the invariant mass distribution, and to about $+2\%$ excess in the low tail of the relative rapidity.
Such minute effects of model parameter variations place extremely high expectations on our theoretical predictions.
 
The top-pair production process is often thought of in terms of Feynman diagrams involving two top quarks, one $t$ and one $\bar{t}$. 
Top quarks are very short lived, however, and the set of final state particles obtained after both top quarks decay can also be realized with diagrams with fewer than two top quarks. 
For example the $W^+W^-b\bar{b}$ final state can also proceed via zero or one top quark.
The production mode with two top quarks, the so-called doubly-resonant production, dominates the other production modes.
However, the ``less-resonant'' modes can match or even surpass the doubly-resonant production in some regions of phase space and are not negligible in the full phase space either.
In particular, the non-doubly-resonant production modes also contribute below the $t\bar{t}$ threshold.
Many modern measurements rely on calculations based on the narrow width approximation~(NWA).
In such calculations the modelling of the non-resonant contribution and its interference with the resonant one is difficult and its reliability may become unsatisfactory as the demand for precision grows.

In these proceedings we investigate the impact of the non-doubly-resonant production mode on the top-pair observables. 
As expected, we find that varying the way how these contributions are modelled only has a mild effect on the shapes of distributions, a few percent at most. 
However, when we then compare such non-resonant effects to those due to the above $Y_t$ variations by eye, we find their magnitudes to be comparable.
Furthermore, contrary to our expectations we also find that the modelling of offshell effects in the top propagator in event generators relying on NWA rapidly fails away from the doubly-resonant region.
It is thus of utmost importance that the best theoretical models of non-resonant and offshell effects available are considered.
Note that the results presented in these proceedings are original.

\section{Calculation and setup}
The NLO {\tt POWHEG\,BOX}~\cite{Nason:2004rx,Frixione:2007vw,Alioli:2010xd,Jezo:2015aia} generators that we used in this study to simulate the process $pp\to b\bar{b}e^+\mu^-\nu_e \bar{\nu}_\mu$, which is dominated by top-pair production with subsequent fully leptonic decay, are the following:
\begin{itemize}
    \item ``bb4l''~\cite{Jezo:2016ujg}, which describes the process at NLO QCD, including offshell effects and the non-resonant $tW$ contribution exactly;
    \item ``ttdec''~\cite{Campbell:2014kua}, which describes the process exactly at LO. 
    QCD radiative corrections are implemented in the NWA only for the $t\bar{t}$ double-resonant channel;
    \item ``ttdec dr''~\cite{Campbell:2014kua}, which describes $pp\to t\bar{t}$, including the top decay in NWA, at NLO QCD. LO finite top-width effects are included exactly via reweighting. 
    \item ``hvq''~\cite{Frixione:2007nw}, which describes the process $pp\to t\bar{t}$ at NLO QCD. The top decay is implemented at LO via a reweighting technique which allows one to partially take into account spin correlations and offshell effects. This generator is used for the modelling of the doubly-resonant top-pair production in the CMS $Y_t$ extraction of Ref.~\cite{Sirunyan:2020eds}.
\end{itemize}
The ``hvq'' generator does not include the single resonant $tW$ contribution, for this reason one needs to include this channel separately by means of the ``wtch'' generator~\cite{Re:2010bp}. 
However, since the two contributions interfere, one needs to remove the double counting. 
This procedure is somehow arbitrary and can be performed via a diagram removal (DR) or a diagram subtraction (DS) procedure.
We only show results obtained using the DS prescription as the difference between the two prescriptions are typically smaller than the size of the effects investigated here.

The NLO events that we generate are showered with the {\tt Pythia 8.244} parton shower~\cite{Sjostrand:2014zea}, which includes matrix element corrections to correct the hardest emission from the top decay whether or not it is provided by the {\tt POWHEG} generator. 
We do not include QED radiation and non-perturbative physics effects. 

We simulate LHC events at $\sqrt{s}=13$~TeV, using the {\tt NNPDF3.1\_nlo\_as\_118} set of parton distribution functions \cite{Ball:2017nwa}. 
We set $m_t=172.5$~GeV, $m_b=4.75$~GeV and we choose as the central renormalization and factorization scale $\mu_R=\mu_F=m_t$. 
We consider a fiducial phase space in which event selection is based on the following criteria:
\begin{itemize}
    \item at least one $e^+$ and one $\mu^-$ with $p_{\perp\ell} > 20$~GeV and $|\eta_\ell|<2.4$, the hardest lepton must also satisfy $p_{\perp\ell^{\rm hard}}>30$~GeV;
    \item for the missing transverse momentum we require $p_{\perp}^{\min}>30$~GeV;
    \item two $b$-jets (clustered with the {\tt FastJet} implementation of the anti-$k_t$ algorithm with $R=0.4$, \cite{Cacciari:2011ma}) with $p_{\perp b}>30$~GeV and $|\eta_b|<$2.4. We denote the jet containing the $b$ ($\bar{b}$) quark with $j_{b}$ ($j_{\bar{b}}$).
\end{itemize}
We pair leptons and its corresponding $b$-jets at the Monte Carlo truth level.

\section{Results}
In this section we present our predictions for the invariant mass of the $e^+\mu^-j_b j_{\bar{b}}$ system, $m(e^+\mu^-j_b j_{\bar{b}})$, and for the rapidity separation of the $e^+j_b$ and $\mu^-j_{\bar{b}}$ systems, $\Delta y(e^+j_b,\mu^-j_{\bar{b}})$, that we obtain using the various top-pair and $tW$ event generators available in {\tt POWHEG\,BOX}.
We compare the results at leading order (LO) and at next-to-leading order matched to parton shower (NLO+PS) level, in the full and in the fiducial phase space as defined in the previous section.

We interpret the observed differences in terms of two effects: the modelling of the top-quark offshellness and of the non-doubly-resonant contributions including the quantum interference with the doubly-resonant one.
At LO we expect the ``ttdec'' and ``bb4l'' generators to perform best, ``ttdec dr'' to be lacking non-double-resonant contributions and ``hvq'' to be similar to ``ttdec dr'', up to subleading differences in the modelling of the top propagator.
At NLO+PS we expect the ``bb4l'' generator to perform best, ``ttdec'' to differ due to missing higher order corrections to the non-resonant contributions, and the combination of ``hvq'' and ``wtch'' generators to differ due its intrinsic ambiguity in the treatment of the resonant--non-resonant interference effects. 

\begin{figure}[tb]
    \centering
    \includegraphics[width=0.40\textwidth]{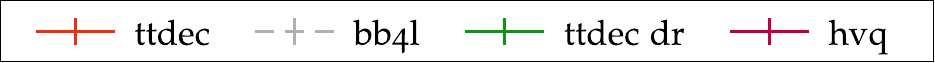}\\
    \vskip0.3em
    \includegraphics[width=0.495\textwidth]{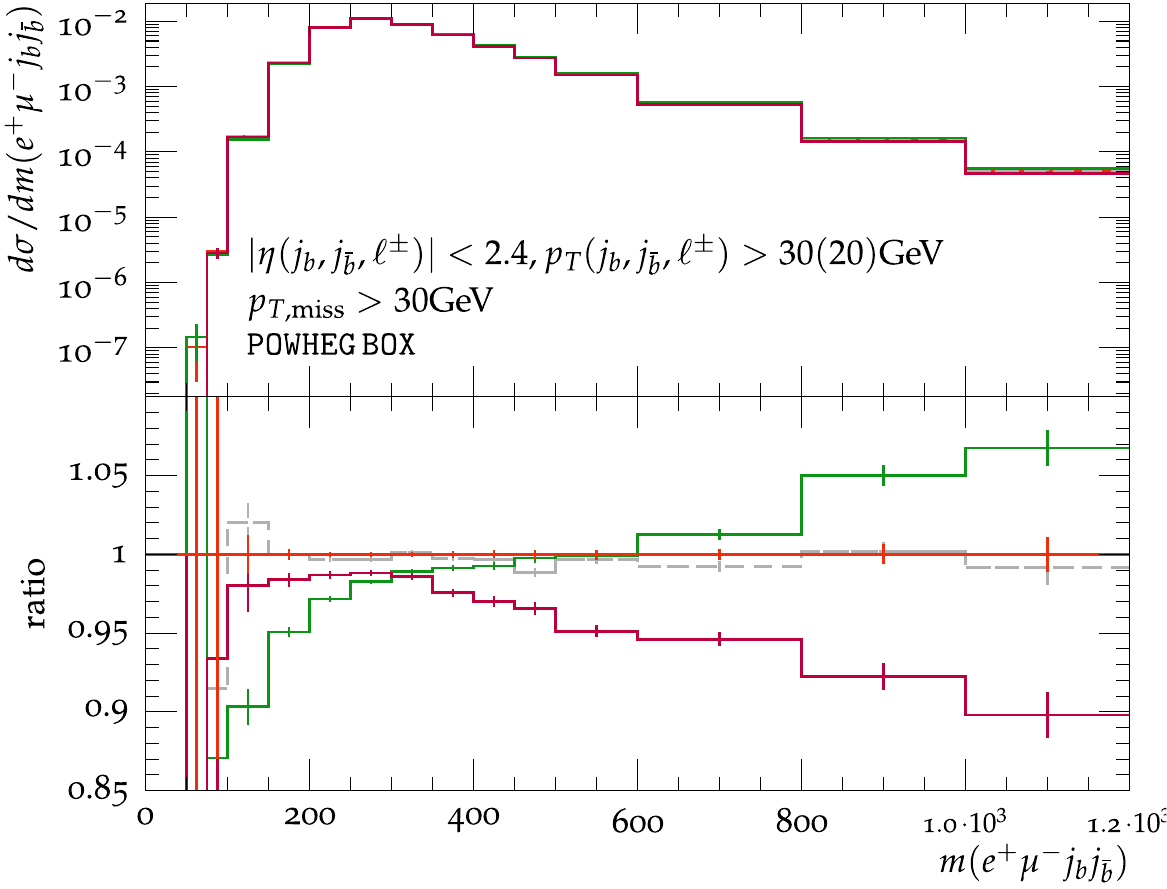}
    \includegraphics[width=0.495\textwidth]{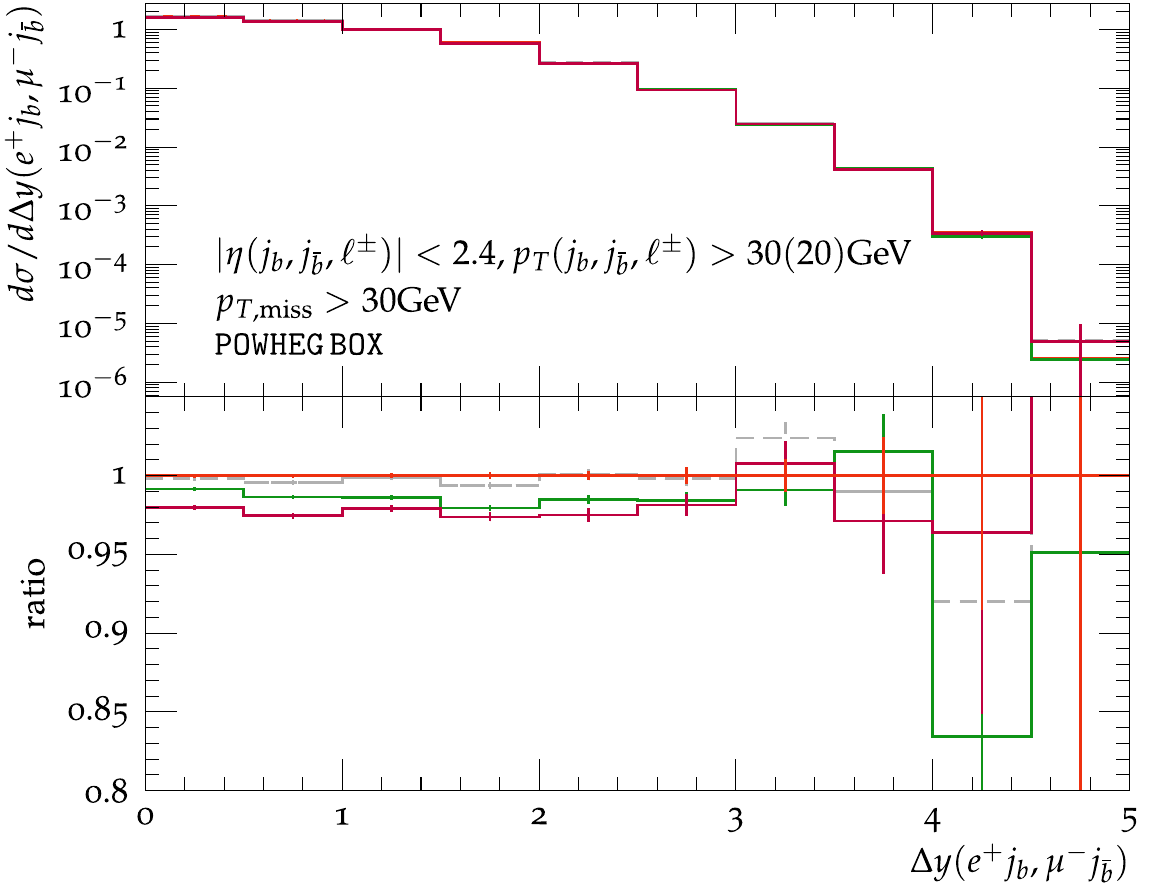}
    \caption{LO predictions for $m(e^+\mu^-j_b j_{\bar{b}})$ (left) and for $\Delta y(e^+j_b, \mu^-j_{\bar{b}})$ (right) in the upper panels. The lower panels show the ratios of the predictions in the upper panels to ``ttdec''. }
    \label{fig:leadingOrderDistributions}
\end{figure}

In Fig.~\ref{fig:leadingOrderDistributions},
we present LO predictions for $m(e^+\mu^-j_b j_{\bar{b}})$ (left) and for $\Delta y(e^+j_b, \mu^-j_{\bar{b}})$ (right). 
We show the predictions from ``ttdec'' in red, ``ttdec dr'' in green and ``hvq'' in purple.
In the lower panels of both plots  the ratios of those predictions with respect to our best prediction from ``ttdec'', which we choose as reference, are displayed.
We find that across the peaks of $m(e^+\mu^-j_b j_{\bar{b}})$ all three predictions agree within a couple of percent when normalized to the same cross sections.
Outside of this range the non-doubly-resonant contributions increase the cross section in the lower ($< 200$ GeV) and decrease it in the higher tail ($> 800$ GeV) by almost 10\%, which is from comparing the ``ttdec dr'' and the ``ttdec'' curves.
The ``hvq'' generator, similarly to ``ttdec dr'', also omits the non-resonant contributions and so one would expect the two to agree well.
Instead, we find that that careful modelling of the top propagator is as important for $m(e^+\mu^-j_b j_{\bar{b}})$ as the inclusion of non-resonant production modes, if not more. 
This is not completely unexpected, as in Ref.~\cite{FerrarioRavasio:2019vmq} it was already shown that the ``hvq'' generator is inadequate to describe observables sensitive to correlations between the decay products of the two top quarks.
Overall the ``hvq'' prediction, in NWA and with approximate top decay modelling, undershoots the fully-offshell ``ttdec'' prediction both in the low and the high tail by $\sim 5\%$ and $\sim 10\%$ respectively.
All three predictions for $\Delta y(e^+j_b, \mu^-j_{\bar{b}})$ agree extremely well shape-wise.

Our findings for the invariant mass distributions do not depend appreciably on whether the fiducial or the inclusive phase spaces are considered. 
This is not the case for the rapidity separation distribution, for which shape effects of about 10\% present originally present in the inclusive phase space completely disappear in the fiducial phase space.
At LO the ``ttdec'' and ``bb4l'' (dashed gray) event generators should reproduce one another. 
Indeed, we find an excellent agreement.

\begin{figure}[tb]
    \centering
    \includegraphics[width=0.40\textwidth]{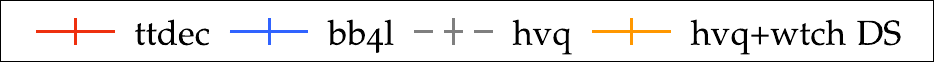}\\
    \vskip0.3em
    \includegraphics[width=0.495\textwidth]{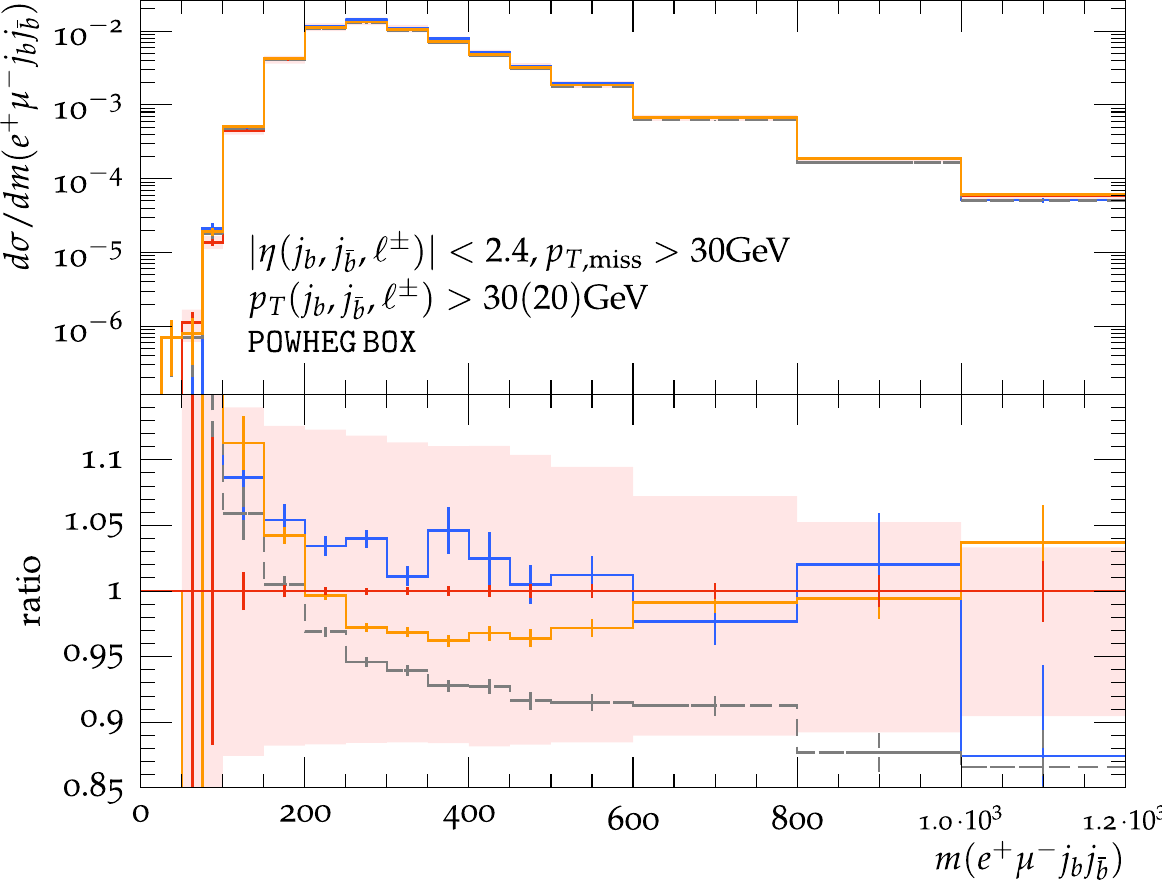}
    \includegraphics[width=0.495\textwidth]{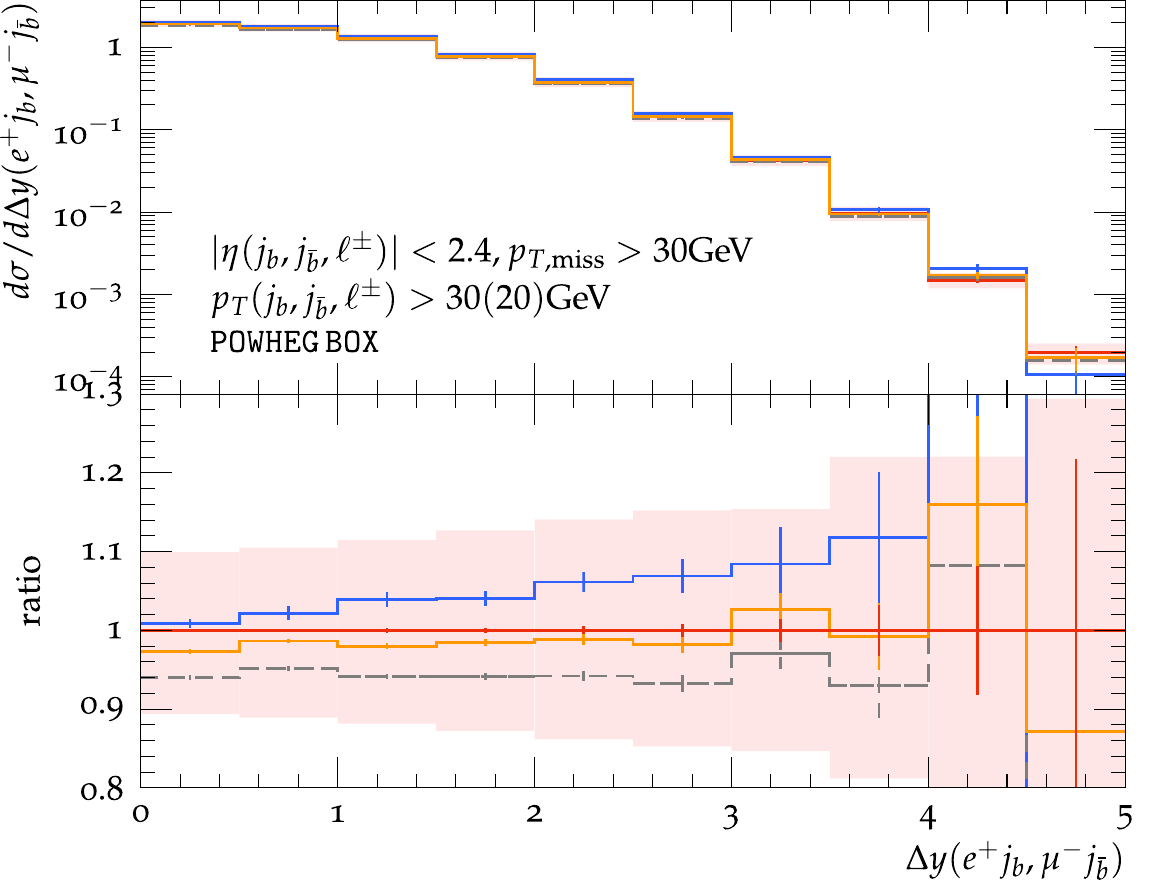}
    \caption{NLO+PS predictions for $m(e^+\mu^-j_b j_{\bar{b}})$ (left) and for $\Delta y(e^+j_b, \mu^-j_{\bar{b}})$ (right) in the upper panels. The lower panels show the ratios of the predictions in the upper panels to ``ttdec''.}
    \label{fig:nextToLeadingOrderDistributions}
\end{figure}

In Fig.~\ref{fig:nextToLeadingOrderDistributions}
we display the $m(e^+\mu^-j_b j_{\bar{b}})$ (left) and the $\Delta y(e^+j_b, \mu^-j_{\bar{b}})$ (right) distributions as before, but this time at NLO+PS.
We show the predictions from ``ttdec'' in red including the error band due to the usual 7-point renormalization and factorization scale variation, ``bb4l'' in blue and ``hvq+wtch'' in orange.
The ratios of those predictions with respect to the ``ttdec'' are displayed in the lower panels. 

While at the LO all the predictions for $\Delta y(e^+j_b, \mu^-j_{\bar{b}})$ agree very well in the fiducial phase space, this is no longer the case at NLO+PS. 
As before, the ``ttdec'' and ``hvq+wtch'' samples have the same shape, while ``bb4l'' decreases appreciably slower reaching difference of about 10\% at $\Delta y = 4$.

In the fiducial phase space and in the range of $m(e^+\mu^-j_b j_{\bar{b}})$ above 150 GeV both the ``ttdec'' and ``hvq+wtch'' samples are at most 5\% and 10\% away from the ``bb4l'' sample and the shape of ``bb4l'' prediction is a bit closer to the ``ttdec'' than to ``hvq+wtch''.
In the full phase space, which we do not show here, the higher level of agreement between ``ttdec'' vs. ``bb4l'' relative to ``hvq+wtch'' vs. ``bb4l'' is more evident. 
This may suggest that the higher order QCD corrections to the non-resonant production and its interference with the doubly-resonant production, missing in ``ttdec'', have less of a shape effect than the mis-modelling of the whole interference term in ``hvq+wtch''.
Nevertheless, all generators agree within the ``ttdec'' perturbative scale uncertainty both in the full and the fiducial phase space, with the exception of the very high tail of the ``hvq+wtch'' sample in the full phase space (not shown here).

In order to illustrate the impact of the ``wtch'' contribution to ``hvq+wtch'', we also show the ``hvq'' prediction in dashed gray. 
The ``wtch'' prediction for $m(e^+\mu^-j_b j_{\bar{b}})$ has quite a similar shape as ``hvq'' below 400 GeV the but grows appreciably faster than ``hvq'' above.

\section{Summary}
We investigate the impact of non-resonant production on traditional top-pair observables $m(e^+\mu^-j_b j_{\bar{b}})$ and $\Delta y(e^+j_b, \mu^-j_{\bar{b}})$, often used for measurements of fundamental top quark parameters such as $Y_t$ or the top mass.
We find that the inclusion of the non-doubly-resonant contribution as well as the modelling of its interference with the resonant production, which are both very important not only near the top-pair threshold, can modify the shape of both distributions with a similar magnitude as small variations of the top parameters that we want to infer.
We also observe that the mis-modelling of the offshell effects in the top propagator can lead to substantial shape distortions.
It is thus of extreme importance to take such effects into account. 

\Acknowledgements
We are grateful to Michelangelo Mangano for the idea for this study and for his valuable feedback on this manuscript. 
We would also like to thank Paolo Nason for his guidance and useful discussions.
We thank Katie Whitfield for careful proofreading.

\end{document}

%% file: econfmacros.tex
\def\beq{\begin{equation}}
\def\eeq#1{\label{#1}\end{equation}}
\def\eeqn{\end{equation}}

\def\beqa{\begin{eqnarray}}
\def\eeqa#1{\label{#1}\end{eqnarray}}
\def\eeqan{\end{eqnarray}}

\let\bar=\overbar

\def\Dslash{\not{\hbox{\kern-4pt $D$}}}
\def\dslash{\not{\hbox{\kern-2pt $\del$}}}

\def\msb{{\bar{\ssstyle M \kern -1pt S}}}

%% file: eprint.bbl
\begin{thebibliography}{99}

\bibitem{Aaboud:2018urx}
M.~Aaboud \textit{et al.} [ATLAS],
%``Observation of Higgs boson production in association with a top quark pair at the LHC with the ATLAS detector,''
Phys. Lett. B \textbf{784}, 173-191 (2018)

\bibitem{Sirunyan:2018hoz}
A.~M.~Sirunyan \textit{et al.} [CMS],
%``Observation of $\mathrm{t\overline{t}}$H production,''
Phys. Rev. Lett. \textbf{120}, no.23, 231801 (2018)

\bibitem{Sirunyan:2019nlw}
A.M.~Sirunyan et al.,
CMS Collaboration, 
%Measurement of the top quark Yukawa coupling from $\mathrm{t\bar{t}}$ kinematic distributions in the lepton+jets final state in proton-proton collisions at $\sqrt{s} =$ 13 TeV
Phys.Rev. \textbf{D100} (2019) 072007

\bibitem{Sirunyan:2020eds}
A.M.~Sirunyan et al.,
CMS Collaboration, 
%Measurement of the top quark Yukawa coupling from $\mathrm{t\bar{t}}$ kinematic distributions in the dilepton final state in proton-proton collisions at $\sqrt{s}=$ 13 TeV
Phys.Rev. \textbf{D102} (2020) 092013

\bibitem{Nason:2004rx}
P.~Nason,
%``A New method for combining NLO QCD with shower Monte Carlo algorithms,''
JHEP \textbf{11} (2004), 040

\bibitem{Frixione:2007vw}
S.~Frixione, P.~Nason and C.~Oleari,
%``Matching NLO QCD computations with Parton Shower simulations: the POWHEG method,''
JHEP \textbf{11} (2007), 070

\bibitem{Alioli:2010xd}
S.~Alioli, P.~Nason, C.~Oleari and E.~Re,
%``A general framework for implementing NLO calculations in shower Monte Carlo programs: the POWHEG BOX,''
JHEP \textbf{06} (2010), 043

\bibitem{Jezo:2015aia}
T.~Je\v{z}o and P.~Nason,
%``On the Treatment of Resonances in Next-to-Leading Order Calculations Matched to a Parton Shower,''
JHEP \textbf{12} (2015), 065

\bibitem{Jezo:2016ujg}
T.~Je\v{z}o, J.~M.~Lindert, P.~Nason, C.~Oleari and S.~Pozzorini,
%``An NLO+PS generator for $t\bar{t}$ and $Wt$ production and decay including non-resonant and interference effects,''
Eur. Phys. J. C \textbf{76} (2016) no.12, 691

\bibitem{Campbell:2014kua}
J.~M.~Campbell, R.~K.~Ellis, P.~Nason and E.~Re,
%``Top-Pair Production and Decay at NLO Matched with Parton Showers,''
JHEP \textbf{04} (2015), 114

\bibitem{Frixione:2007nw}
S.~Frixione, P.~Nason and G.~Ridolfi,
%``A Positive-weight next-to-leading-order Monte Carlo for heavy flavour hadroproduction,''
JHEP \textbf{09} (2007), 126

\bibitem{Re:2010bp}
E.~Re,
%``Single-top Wt-channel production matched with parton showers using the POWHEG method,''
Eur. Phys. J. C \textbf{71} (2011), 1547

\bibitem{Sjostrand:2014zea}
T.~Sj\"ostrand, S.~Ask, J.~R.~Christiansen, R.~Corke, N.~Desai, P.~Ilten, S.~Mrenna, S.~Prestel, C.~O.~Rasmussen and P.~Z.~Skands,
%``An introduction to PYTHIA 8.2,''
Comput. Phys. Commun. \textbf{191} (2015), 159-177

\bibitem{Ball:2017nwa}
R.~D.~Ball \textit{et al.} [NNPDF],
%``Parton distributions from high-precision collider data,''
Eur. Phys. J. C \textbf{77} (2017) no.10, 663

\bibitem{Cacciari:2011ma}
M.~Cacciari, G.~P.~Salam and G.~Soyez,
%``FastJet User Manual,''
Eur. Phys. J. C \textbf{72} (2012), 1896

%\cite{FerrarioRavasio:2019vmq}
\bibitem{FerrarioRavasio:2019vmq}
S.~Ferrario Ravasio, T.~Je\v{z}o, P.~Nason and C.~Oleari,
%``A theoretical study of top-mass measurements at the LHC using NLO+PS generators of increasing accuracy,''
Eur. Phys. J. C \textbf{78} (2018) no.6, 458


\end{thebibliography}
